# Proportional Topology Optimization: A new non-gradient method for solving stress constrained and minimum compliance problems and its implementation in MATLAB

Emre Biyikli, Albert C. To

Department of Mechanical Engineering and Materials Science & Center of Simulation and Modeling, University of Pittsburgh, Pittsburgh, PA, USA

A new topology optimization method called the Proportional Topology Optimization (PTO) is presented. As a non-gradient method, PTO is simple to understand, easy to implement, and is also efficient and accurate at the same time. It is implemented into two MATLAB programs to solve the stress constrained and minimum compliance problems. Descriptions of the algorithm and computer programs are provided in detail. The method is applied to solve three numerical examples for both types of problems. The method shows comparable efficiency and accuracy with an existing gradient optimality criteria method. Also, the PTO stress constrained algorithm and minimum compliance algorithm are compared by feeding output from one algorithm to the other in an alternative manner, where the former yields lower maximum stress and volume fraction but higher compliance compared to the latter. Advantages and disadvantages of the proposed method and future works are discussed. The computer programs are self-contained and publicly shared in the website www.ptomethod.org.

**Corresponding author:** Albert C. To. Address: 508 Benedum Hall, University of Pittsburgh, PA 15261, USA. Tel.: +1 (412) 624 2052. e-mail address: albertto@pitt.edu.

**Introduction**

Topology optimization can be viewed as the systematic removal of redundant material from the design domain in order to attain design with higher strength-to-weight ratios. It is getting an increasing amount of attention since its introduction to truss structures by Michell (1904) and continuum structures by Bendsoe and Kikuchi (1988). Even further, recently popular additive manufacturing techniques appreciate the importance of topology optimization since it facilitates the manufacture of porous structural designs with much complicated geometries.

Topology optimization methods are required to provide designers with black-and-white (or 1/0) designs to easily identify structural members as black regions and voided regions as white regions. On the contrary, it was noticed that topology optimization methods with continuous design variables are more successful for minimization of the objective function (Svanberg and Werme 2007). For this reason, continuum design variables with penalization methods are highly favored, such as the Solid Isotropic Material with Penalization (SIMP) introduced by Bendsoe (1989). It is important to realize that, discrete or continuous, topology optimization is only a conceptual tool and requires post-processing of the optimized geometry. Two popular design problems are the stress constrained problem, which aims at minimizing volume fraction while satisfying stress constraints and the minimum compliance problem, which aims at minimizing compliance for a given volume fraction. In short, these problems will be referred to as stress problem and compliance problem hereafter. Also, the word "element" always refers to the finite element (FE) of an FE mesh in this work. In this context, the design variable can be imagined as thickness of a plate (Bendsoe and Sigmund 2003) or scaling factor of a unit cell in a cellular structure (Zhang et al. 2014).

The compliance problem has been widely investigated by Bendsoe and Sigmund (2003), Sigmund (2001), and Stolpe and Svanberg (2001), to name a few. Open source computer programs to solve this type of problem are distributed (Andreassen et al. 2011; Challis 2010; Liu and Tovar 2014; Sigmund 2001). On the other hand, it is a well-known fact that stress analysis is a more significant concern for designers. Compared to compliance problems, however, stress problems bear more challenging difficulties such as high non-linearity (Le et al. 2010). The stress problem and related issues has been studied by Lee (2012), Duysinx and Bendsoe (1998), and Paris et al. (2009) to name a few. Nevertheless, probably due to its added commercial value and complexity, there is no open source distribution of such a computer program for continua.

Numerous topology optimization techniques have been developed to solve both types of problems. Among these are optimality criteria method, convex linearization method, method of moving asymptotes, successive linear programming, and evolutionary structural optimization method. For a broader list of methods, see Sigmund (2011) and Rozvany (2009).

The Optimality Criteria (OC) method is the most fundamental as compared to the other listed methods (Vemaganti and Lawrence 2005) and was first introduced in structural design by Prager (1968). The method assigns design variables to elements proportionally to the values of the objective function (Bendsoe 1995). In this respect, it is an efficient and simple method. Sigmund et al. (2001) employs the OC method in the TOP99 computer program, which is a 99-line MATLAB code that solves the compliance problem for the Messerschmitt-Bölkow-Blohm (MBB) beam.

The Successive Linear Programming (SLP) method linearizes the originally nonlinear problem at a design point and then locally optimizes the linear problem within a region bounded by some move limits. The local optimization problem can be solved by, for instance, the simplex algorithm (Dantzig 1963). The SQP method is only different from the SLP method in converting the originally nonlinear problem into a quadratic problem. As opposed to SLP, the Convex Linearization (CONLIN) method performs linearization with different variables with respect to the characteristics of the optimization problem (Fleury and Braibant 1986). In this respect, the Method of Moving Asymptotes (MMA) is a specific version of CONLIN in that the search behavior is more aggressively controlled by moving limits (Svanberg 1987). The reader is referred to the book by Christensen (2009) for further details on these methods.

The Evolutionary Structural Optimization (ESO) method starts with a full design domain and then iteratively removes elements from the domain with respect to the values of the objective function (Huang and Xie 2010; Xie and Steven 1992). If the method also includes addition of elements, it is then called Bidirectional ESO (BESO). Addition/removal of elements render a discreteness that is earlier noted as a bad attribute in terms of minimization performance. Indeed, ESO resembles the Fully Stressed Design (FSD) method, which dictates removal of material from an element until the element is fully stressed (Haftka 1992). FSD may also be considered a simple OC method. Its performance to yield an optimal solution is questioned by Rozvany (2009).

Among the introduced methods, SLP, SQP, CONLIN, and MMA require calculation of the gradients of objective function and constraints. OC methods do not necessarily require gradients. In the TOP99 MATLAB code; however, some sort of gradient information is utilized (Sigmund 2001). More specifically, the displacements are held constant and the stiffness matrix is updated by the derivative of the SIMP expression in order to obtain the sensitivity of compliance. On the contrary, more rigorous gradient calculations are usually employed in stress problems (Holmberg et al. 2013; París et al. 2010). These gradients, especially for stress, are analytically complicated and their computation brings an additional computational burden (Patel et al. 2008). Besides, computation of gradients may introduce some implementation concerns (París et al. 2010).

Due to these issues, many non-gradient methods have been cited by Sigmund (2011). In this paper, the usefulness of non-gradient methods is discussed in detail. It is important to note that gradient information is useful to speed up the optimization algorithm. This is proven by many non-gradient methods that cannot show as efficient results as gradient methods, especially the ones based on random processes such as genetic algorithms. Nevertheless, non-gradient methods with comparable efficiencies have also been reported (Sigmund 2011). In short, there is a trade-off between the gradient and non-gradient methods in terms of computational/implementation complexity and efficiency.

In this paper, a simple and efficient non-gradient method, called the Proportional Topology Optimization (PTO), is presented to perform topology optimization for stress (PTOs) and compliance (PTOc) problems. The PTO algorithm assigns the design variables to elements proportionally to the value of stress in the stress problem and compliance in the compliance problem. In particular, it imposes constraints only globally on the entire system. Accordingly, it globally manages the proportional distribution of design variables to the elements. In view of its algorithm, the method can be classified as an OC method. It is admitted that PTO method is highly heuristic and searches for the optimized solutions. Nevertheless, it is this heuristic that makes the method simple to understand and implement. Also, the method does not incorporate gradients; therefore, it avoids the complications associated with gradients. Employment of continuous density variables improves the search performance of the method and preserves the flexibility to design for intermediate densities. Results indicate that the method produces efficient and accurate solutions in consideration of its simplicity.

Inspired by the TOP99 computer program, the method is implemented into two MATLAB programs individually for the stress and compliance problems that solve the MBB beam example. The computer programs are implemented as self-contained MATLAB functions such that they do not even depend on optional MATLAB toolboxes. The authors are distributing the source of computer programs freely for educational and research purposes in the website www.ptomethod.org. To the best of the authors' knowledge, PTOs is the first publicly shared and self-contained computer program that solves the stress constrained problem for continua.

The paper presents, in order, stress and compliance problems, the algorithm, numerical examples, and conclusions. Computer programs are in Appendices A and B.

**Stress constrained problem**

The stress problem is the minimization of volume fraction while satisfying the stress constraints. The optimization problem reads

$$\begin{cases} \min \sum_{i}^{N} \rho_i v_i \\ such\ that \begin{cases} Ku = f \\ \sigma_i \leq \sigma_l\ if\ \rho > 0 \\ 0 \leq \rho_{min} \leq \rho_i \leq \rho_{max} \leq 1 \end{cases} \end{cases} \quad (1)$$

where $N$ is the number of elements, $\rho$ is the density (and also the design variable), $\rho_i$ is the elemental density, $v_i$ is the elemental area/volume, $K$ is the stiffness, $u$ is the displacement, $f$ is the external force, $\sigma_i$ is the elemental stress measure, $\sigma_l$ is the stress limit, $\rho_{min}$ is the lower bound on elemental density, and $\rho_{max}$ is the upper bound on elemental density. Typically, $\rho$ is limited to $[\rho_{min}, 1]$ where $\rho_{min}$ is 0.001 (París et al. 2009) to preclude stiffness singularities (Bruggi 2008). Although the problem is posed as minimization of the total mass, it is usually referred to as minimization of the volume fraction for practical reasons. Minimization of these terms is equivalent from the optimization point of view. A volume fraction 0 means void while 1 means solid element. The stress problem is noted to be non-convex and highly non-linear (París et al. 2009).

**Minimum compliance problem**

The compliance problem is minimization of the compliance while satisfying the volume fraction constraint. The optimization problem reads

$$\begin{cases} \min C = u^T K u \\ such\ that \begin{cases} Ku = f \\ \sum_{i}^{N} \rho_i v_i = M \\ 0 \leq \rho_{min} \leq \rho_i \leq \rho_{max} \leq 1 \end{cases} \end{cases} \quad (2)$$

where, in addition to the nomenclature given for the stress problem, $C$ is the compliance and $M$ is the total mass.

**The PTO algorithms**

Algorithms of the PTO method to solve the stress (PTOs) and compliance (PTOc) problems are described in Figures 1 and 2, respectively.

```
Algorithm
- Setup FE and stress analyses and filtering
- Until convergence
  o Perform FE and stress analyses
  o Check stop criteria, break if satisfied
  o Run optimization algorithm
    ▪ Determine TM
      • Distribute RM
        o If stress limit is exceeded,
          TM = CM + MM
        o Else, TM = CM - MM
    ▪ Set RM = TM
    ▪ Until RM is small enough
      • Distribute RM to elements proportionally
        to their stress values
      • Apply filter
      • Apply density limits
      • Calculate AM
      • Update RM = TM – AM
    ▪ Update density
```

where TM is the target material amount, CM is the current material amount, MM is the material move amount, RM is the remaining material amount, and AM is the actual material amount.

**Figure 1:** PTOs algorithm to solve the stress problem.

Figure 1 presents the PTOs algorithm. The algorithm starts with setup of vectors and matrices for FE and stress analyses and filtering. Then, the algorithm enters the main loop. Every iteration of the main loop starts with FE and stress analyses. Following, the termination criteria is checked. That is, whether the maximum elemental stress in the system is close to the allowable stress limit within a prescribed tolerance, which is set equal to 0.001 in this work. If the criterion returns true, the simulation terminates. Otherwise, the algorithm continues to optimize the topology. The first step of optimization part is to determine the target material amount, which is going to be the new material amount in the system. In other words, the current material amount will be updated to the target material amount. If the maximum elemental stress in the system is bigger than the allowable stress limit, then the current material amount is increased by a material move amount. Otherwise, the current material amount is decreased by the same material move amount. The material move amount scales with the number of elements (0.001 x *number of elements*) and is kept constant during the course of the simulation. In the next step, the algorithm distributes the target material amount to the elements. The target material amount can only be distributed iteratively for the reasons that will be explained in the following. Because of this iterative procedure, the material amount to be distributed is called the remaining material amount, and the iterative procedure initiates with a remaining material amount that is equal to the target material amount.

In order to perform the iterative distribution of target material amount, the algorithm goes into an inner loop. The distribution is conducted proportionally to the elemental stress values. The degree of proportion is extended to the power of $q$ such that

$$\rho_i^{opt} = \frac{RM}{\sum_i^N \sigma_i^q} \sigma_i^q \qquad (3)$$

where $RM$ is the remaining material amount, $N$ is the number of elements, $\rho_i^{opt}$ is the optimized elemental density, $\sigma_i$ is the elemental stress measure, and $q$ is the proportion exponent. Apparently, the above relation distributes the remaining material amount regardless of density limits. The enforcement of density limits on the elements trims the distributed material amount to the lower and upper bounds if the bounds are exceeded. As a result, the actual material amount is different than the target material amount. This difference is the reason for distributing the remaining material amount iteratively in an inner loop until the target material amount is reached. Every iteration of the inner loop starts with distributing the remaining material amount. It is followed by application of filtering and density limits. In this work, a volume preserving density filtering is used, which will be explained in detail later. At the end of the inner loop, the actual material amount, which is left after enforcing limits and filtering, is calculated. The remaining material amount is then the actual material amount subtracted from the target material amount. In the next iteration of inner loop, this remaining material amount is redistributed following the same routine. The inner loop runs until the remaining material amount is small enough.

The final step of main loop updates the elemental densities by linearly blending elemental densities from the previous iteration and optimized elemental densities in the current iteration. The update scheme reads

$$\rho_i^{new} = \alpha \rho_i^{prev} + (1 - \alpha) \rho_i^{opt} \qquad (4)$$

where $\rho_i$ is the elemental density, $\rho^{new}$ is the new elemental density to be passed to the next iteration, $\rho^{prev}$ is the elemental density from the previous iteration, $\rho^{opt}$ is the optimized elemental density in the current iteration, and $\alpha$ is the history coefficient. The history coefficient decides the ratios of elemental densities from both sides. For instance, a value of 0 eliminates elemental density from the previous iteration and indicates no dependence on the history.

```
Algorithm
- Setup FE and compliance analyses and filtering
- Determine TM
- Until convergence
  o Perform FE and compliance analyses
  o Check stop criteria, break if satisfied
  o Run optimization algorithm
    ▪ Set RM = TM
    ▪ Until RM is small enough
      • Distribute RM to elements proportionally
        to their compliance values
      • Apply filter
      • Apply density limits
      • Calculate AM
      • Update RM = TM – AM
    ▪ Update density
```
where TM is the target material amount, CM is the current material amount, MM is the material move amount, RM is the remaining material amount, and AM is the actual material amount.

**Figure 2:** PTOc algorithm to solve the compliance problem.

PTOc algorithm is slightly different from the PTOs algorithm. The most prominent difference is the determination of the target material amount. PTOc algorithm does not need to modify the target material amount since it is constrained to a fixed amount by definition of the problem. For this reason, PTOc algorithm calculates the target material amount once at the beginning of the simulation and uses it thereafter. Another difference is the distribution of the target material amount. PTOc distributes the target material amount proportionally to the elemental compliance values instead of the elemental stress values. The distribution equation then reads

$$\rho_i^{opt} = \frac{RM}{\sum_i^N C_i^q} C_i^q \quad (5)$$

where RM is the remaining material amount, N is the number of elements, $\rho_i^{opt}$ is the optimized elemental density, $C_i$ is the elemental compliance value, and $q$ is the proportion exponent. The elemental compliance values are recalculated for every iteration at the beginning of the main loop. The last difference is the termination criterion of the main loop. The main loop stops if the maximum change in elemental densities between two successive iterations is smaller than a prescribed tolerance, which is equal to 0.01 in this work. The rest of the steps are identical to the PTOs algorithm.

**Material model**

PTO method adopts the modified SIMP approach (Andreassen et al. 2011), which is a density approach, for better search performance while maintaining near 0/1 solutions. The modified SIMP approach reads

$$E(\rho) = E_{min} + \rho^p E_0 \quad (6)$$

where $E$ is the density dependent Young's modulus, $E_{min}$ is a small Young's modulus (typically $10^{-9}$) assigned to void elements, $E_0$ is the Young's modulus of the solid material, and $p$ is the penalty coefficient (typically 3). The modified SIMP approach makes it redundant to have a lower bound for density $\rho_{min}$ to avoid the stiffness singularities since $E_{min}$ already serves the said purpose. The modified SIMP approach drives densities towards 0 and 1 since volume varies linearly as stiffness varies in the order of $p$.

**Stress constraint**

PTO method employs the following maximum function as a stress constraint

$$max\{\sigma_i\} \leq \sigma_{elastic\ limit} \quad (7)$$

where $\sigma_i$ is the stress at element $i$ and it is taken to be the von Mises stress at the geometric center of the element. The details of stress calculation are presented in the following. The stress constraint entails that the stress does not exceed the elastic limit at any element in the system. Therefore, the constraint provides a tight control on the stress levels owing to the maximum function. It should be noted that the maximum function is not differentiable, and thus cannot be used with gradient methods. Instead, gradient methods usually employ a $p$-norm of stress (Le et al. 2010). The $p$-norm stress measure is not as tight as the maximum stress measure unless the value of $p$ is very big. As such, for $p = \infty$, the p-norm stress measure is equivalent to the maximum stress measure. In addition, the $p$-norm stress measure does not have a physical meaning as the maximum stress measure does (Le et al. 2010). Finally, implementation of the maximum stress measure is the simplest compared to the other stress measures.

**Density filtering**

The PTO method incorporates a density filtering. In the work of Bruns (2001), a simple cone density filtering is introduced as the following

$$\rho_i = \frac{\sum w_{ij} d_j}{\sum w_{ij}} \ where\ w_{ij} = \begin{cases} \frac{r_0 - r_{ij}}{r_0} & for\ r_{ij} < r_0 \\ 0 & for\ r_{ij} \geq r_0 \end{cases} \quad (8)$$

$\rho_i$ is the filtered density of element $i$, $w_{ij}$ is the filtering weight of elements $i$ and $j$, $d_j$ is the non-filtered density of element $j$, $r_{ij}$ is the distance between elements $i$ and $j$, and $r_0$ is the filter radius. The weight is inversely proportional to the distance between the element and its neighbors. In this sense, the cone density filtering is actually nothing but local averaging. Besides, it preserves the volume. It should be noted that it is always filtered densities that are presented in the results section. Filtering is endorsed to be advantageous for many reasons:

(i) Small scale features such as jagged edges, narrow members, and sharp interfaces are prevented (Le et al. 2010).

(ii) As a result of smoothing, a blurred region around the structural members is obtained (Le et al. 2010).

(ii) The algorithm is saved from getting stuck in local minima (Le et al. 2010).

(iv) Checkerboard phenomenon is prevented (Sigmund 2007).

(v) Ensures existence of solution, although this is not proven yet (Sigmund 2001).

(vi) Imposes a constraint on minimum length scale of the design (Sigmund 2007).

As a separate note, even if the method had sensitivity, it is argued that sensitivity filtering is not suitable for the stress problem (Le et al. 2010). A number of filtering methods are presented by Sigmund (2007). In addition, two alternative filtering schemes for the Top88 code are introduced by Andreassen (2011).

**Control parameters**

Two control parameters are defined to fine tune the behavior of the PTO algorithm: proportion exponent (q) and history coefficient (α). The proportion exponent controls the degree of proportion between the elemental density value and elemental stress or compliance values for the stress and compliance problems, respectively. For instance, a quadratic proportion for the stress problem means that the total material amount is distributed to elements in proportion to the square of the elemental stress values. The other control parameter is the history coefficient. It controls the ratio of dependence of elemental density to its older value from the previous iteration. For instance, a value of 0.5 means that the elemental densities are blended such that half of their new values come from the previous iteration and the other half come from the optimized values in the current iteration.

A preliminary parametric study reveals that the optimum values of proportion exponent are 2.0 for the stress problem and 1.0 for the compliance problem. Thus, the proportion is quadratic for the stress problem and linear for the compliance problem. Since proportion exponent has no effect in the compliance problem, it is omitted from the presented computer program for the compliance problem. The study also reveals that optimum values for the history coefficient are 0.0 for the stress problem and 0.5 for the compliance problem. Since the stress problem does not include any dependence on history, $\alpha$ is omitted from the presented computer program for the stress problem. A more comprehensive parametric study to utilize the method at its best is left for future work.

**Boundary conditions**

Finite element (FE) problem definitions are required to be accompanied with some essential and natural boundary conditions. These prescribed boundary conditions are usually concentrated and their correct imposition to the problem domain is crucial for the FE solution. In a similar manner, it is vital to correctly handle the boundary conditions for the topology optimization solution. We experienced that exclusion of the elements near the boundary conditions from the topology optimization problem actually results with different solutions from those obtained when these elements are included. Moreover, the exclusion of elements near the boundary conditions yields better optimization results, which may be misleading. On the other hand, imposition of boundary conditions to only a few elements leads to poor topology optimization behavior due to compliance/stress concentration (Duysinx and Bendsøe 1998; Le et al. 2010; Pereira et al. 2004). Consequently, the best practice is to distribute the boundary conditions to a sufficient number of elements in order to provide the topology optimization algorithm to work properly, as followed by many researchers (Deqing et al. 2000; Duysinx and Bendsøe 1998; Le et al. 2010). If the resulting structure is suspected to be fragile for loading conditions as pointed out by Holmberg (2013), more material can be added near the loading regions at the post-processing phase.

**Stress measure**

As stated earlier, von Mises stress is measured at the geometric center of the elements. In the following, only two-dimensional (2-D) examples with plane stress and bilinear square elements of length $L$ are considered. The von Mises stress in 2-D is given by

$$\sigma_{vM} = \sqrt{\sigma_x^2 + \sigma_y^2 - \sigma_x \sigma_y + 3\sigma_{xy}^2} \qquad (9)$$

The stress tensor in 2-D is expressed as

$$\sigma = \begin{Bmatrix} \sigma_x \\ \sigma_y \\ \sigma_{xy} \end{Bmatrix} \qquad (10)$$

And obtained by

$$\sigma = DBu \qquad (11)$$

where $D$ is the constitutive matrix, $B$ is the shape function derivative matrix, and $u$ is the displacement vector. The constitutive matrix for plane stress in 2-D is as the following

$$D = \frac{E}{1-v^2} \begin{bmatrix} 1 & v & 0 \\ v & 1 & 0 \\ 0 & 0 & (1-v)/2 \end{bmatrix} \qquad (12)$$

where $E$ is the Young's modulus and $v$ is the Poisson's ratio. For linear shape functions for a bilinear square element in 2-D, $B$ is given by

$$B = \frac{1}{2L} \begin{bmatrix} -1 & 0 & 1 & 0 & 1 & 0 & -1 & 0 \\ 0 & -1 & 0 & -1 & 0 & 1 & 0 & 1 \\ -1 & -1 & -1 & 1 & 1 & 1 & 1 & -1 \end{bmatrix} \qquad (13)$$

Lastly, $u$ is the element displacement vector represented as

$$u = \begin{Bmatrix} u_{1x} \\ u_{1y} \\ u_{2x} \\ u_{2y} \\ u_{3x} \\ u_{3y} \\ u_{4x} \\ u_{4y} \end{Bmatrix} \quad (14)$$

The term "stress" in the results section always refers to the von Mises stress at the geometric center of the square elements.

**MATLAB programs**

Two separate MATLAB programs that solve the stress and compliance problems for the MBB beam in bending (Fig. 3a) are presented. In short, the MBB beam in bending is referred to as MBB beam hereafter. It is important to acknowledge that the computer programs substantially inherit from the 88-line MATLAB code by Andreassen et al. (Top88 hereafter), such as setup and solution of FE system. In particular, the only major modification is undertaken in optimization algorithm and some other minor modifications elsewhere. Minor modifications include addition of stress analysis and removal of sensitivity analysis. Furthermore, a few extra input parameters are introduced to control: the element edge length, number of elements the load is distributed on, and lower and upper bounds on density. The latter is introduced for different design needs as it may be asked to have a lower bound on density for a cellular structure. This intervention should not conflict with the SIMP approach as long as the penalization factor *penal* is accordingly justified.

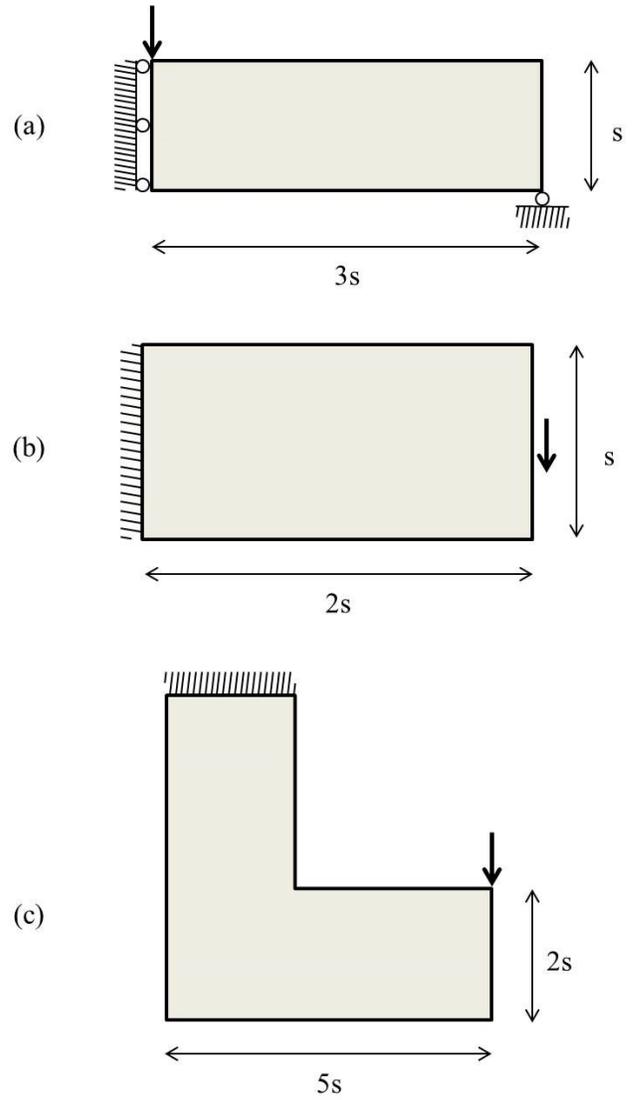

**Figure 3:** Numerical examples: (a) MBB beam – only right half is considered due to symmetry, (b) Cantilever beam, and (c) L bracket.

The computer programs are cast as MATLAB functions that can be called from the MATLAB command window or other MATLAB programs. The first computer program is for the MBB beam example solved for the stress problem (Appendix A). In this case, the function is called as the following

$$x = PTOs\_mbb\ (E0, Emin, L, lv, ld, nelx, nely, nu, penal, q, rmin, vmslim, xlim)$$

where $x$ is the elemental densities, $E_0$ is the Young's modulus, $E_{min}$ is the Young's modulus assigned to void elements, $L$ is the element edge length, $lv$ is the load value, $ld$ is the number of elements displacement and force loads are distributed on, *nelx* is the number of elements in $x$ dimension, *nely* is the number of elements in $y$ dimension, *nu* is the Poisson's ratio, *penal* is the penalization factor in the modified SIMP formula, $q$ is the proportion exponent, *rmin* is the filter radius, *vmslim* is the stress constraint limit, and *xlim* is a *1x2* vector consisting of lower and upper bounds on density, respectively.

Lines 5-9 prepare the element stiffness matrix *KE* that is to be multiplied by the Young's modulus *E* to get to its final form. Lines 10-12 prepare the *edofMat* matrix that is in size of *(element number) x (8)* and consists of degrees of freedoms (DOF) of each element in a row. Numbering of DOF, nodes, and elements in the system starts from top-left and proceeds in column-wise order (Fig. 4).

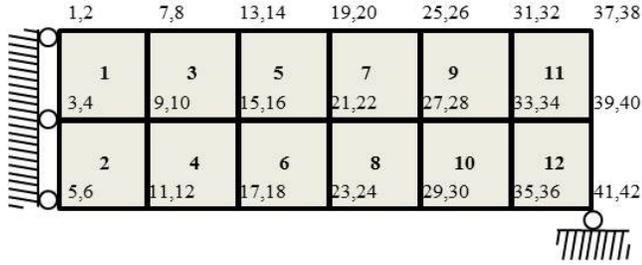

**Figure 4:** Numbering of DOF, nodes, and elements in right half of the MBB beam: starting from top-left and proceeding in column-wise order.

Lines 13-14 prepare *iK* and *jK* vectors that represent the indices of nodes in the global stiffness matrix. Lines 16-19 form the force sparse vector *F* with respect to input load value *lv* and distribution parameter *ld*. Line 21 initializes the displacement vector *U* to zero. Line 22 composes the set of fixed DOFs with respect to the input load distribution parameter *ld*. Lines 23 and 24 composes the sets of all DOFs and free DOFs, respectively. The set of free DOFs *freedofs* is later employed when solving the FE system. Lines 26-27 prepare the element shape function derivative matrix *B* and constitutive matrix *DE* for stress analysis. The latter is to be multiplied by the Young's modulus *E* to get to its final form.

Lines 29-48 build the density filter sparse matrix. In specific, lines 33 and 34 loop for every element position, and lines 36 and 37 loop for neighbor element positions. Lines 40-41 save the indices for pair of neighbors. Line 42 computes and saves the weight of density filtering for the pair of neighbors from the distance between them if the distance is smaller than the input filter radius. After exiting the loop, lines 47 and 48 create the density filter sparse matrix and normalize it, in order.

The main loop takes place between lines 53 and 90. It first carries out the FE analysis in lines 56-59, and finds the displacements *U*. More specifically, the main loop conducts FE analysis by populating the global stiffness sparse matrix *K* in lines 57-58 with the updated Young's modulus values *E* from line 56, and then solving the FE system *KU = f* in line 59. The main loop follows by the stress and compliance analyses. Stress analysis computes the elemental stress tensors in line 61 and the elemental equivalent von Mises stresses in line 62. Compliance analysis computes the elemental compliances into a vector in line 64 and reshapes this vector into a matrix by the corresponding number of elements in each dimension in line 65. The main loop prints out the results to the command window in lines 67-68; and, plots the elemental densities and stresses normalized by the maximum value of corresponding matrices in lines 70-72.

In line 74, the main loop checks for the termination criteria, that is whether the maximum elemental stress in the system is close to the stress constraint limit within a tolerance (i.e., 0.001) and number of iterations is more than 50. The latter is introduced to inhibit immature terminations, which occurred only one time in authors' experience. If the termination criterion returns true, the main loop exits, and simulation ends.

Lines 76-89 consist of the core PTOs algorithm. Initially, lines 76-80 determine the target material amount with respect to the maximum elemental stress in the system. In that, if the maximum elemental stress exceeds the stress constraint, more material is added, or removed otherwise. The added/removed material amount is equal to the multiplication of the total number of elements by 0.001. Following, lines 84-89 represents the inner loop that iteratively distributes the target material amount proportionally to the elemental stress values. This proportion is computed out of the loop in line 83 for sake of efficiency. The proportion is extended by the proportion exponent *q*. The inner loop starts with distribution of the remaining material in line 85. Then, lines 86 and 87 filter the distributed material and enforce density limits on the elemental densities, respectively. The inner loop ends with computation of remaining material amount in line 88. The inner loop terminates when the remaining material amount is less than or equal to 0.001, as checked in line 84.

The second computer program is for the MBB beam example solved for the compliance problem (Appendix B). In this case, the function is called as the following

$$x = PTOc\_mbb\,(alpha, E0, Emin, L, lv, ld, nelx, nely,$$
$$nu, penal, rmin, vmslim, xlim)$$

where *alpha* (i.e., $\alpha$) is the history coefficient and other arguments are identical to PTOs, except that the proportion exponent *q* is omitted. Although the lines of PTOs and PTOc do not match at the same line number all the time, the flow and steps of the programs are largely the same. The differences are detailed in the following.

PTOc has a new variable that first appears in line 51, named *xNew*, and stores the optimized elemental densities in the current iteration of the loop. Later, line 88 updates elemental densities *x* with respect to the history coefficient *alpha* as a linear combination of elemental densities from the previous (i.e., *x*) and current (i.e., *xNew*) iterations. Line 76 checks whether the termination criteria is satisfied. That is, if change in the maximum elemental densities between two successive iterations (this *change* is computed in line 89) is smaller than 0.01 and the number of iterations is more than 50. The former criterion is different than that of PTOs since PTOc satisfies the volume constraint a priori in line 78 as will be explained later. In contrast, PTOs searches for a distribution until the stress constraint is satisfied, hence a posteriori.

Line 78 computes the target material amount as dictated by the input constraint on element volume fraction *vlim*. This value is constant during the course of the simulation. As can be followed from lines 81 and 83, PTOc distributes the material amount in proportion to the elemental compliance values. The proportion is more direct (and linear) compared to PTOs since there is no use of proportion exponent.

In case the above descriptions of computer programs are not clear enough, the user is referred to two other MATLAB codes and corresponding papers, namely 99-line code (Sigmund 2001) and 88-line code (Andreassen et al. 2011), for alternative descriptions due to the fact that current codes mainly inherit from the two referred codes.

The computer programs are highly flexible and extensible. For instance, the programs can easily be modified to insert a prescribed void or solid region in the design by constraining the corresponding elemental densities to 0 or 1 in the inner loop right after updating *x* in line 87 in PTOs and 85 in PTOc. For another instance, PTOs can be extended to minimize volume fraction under both stress and compliance constraints. Then, in addition to the check for elemental stresses, the same practices should be implemented for elemental compliances. This way, material should be added to the system when either of the constraints is not satisfied, and material should be removed from the system when both constraints are satisfied. In like manner, the simulation should terminate when both constraints are satisfied at the same time.

The computer programs are unitless. However, a set of units can be attached to attain a physical relevance. A set of consistent units are kg for mass, meter for length, and second for time. Then, force units are Newton, stress units are Pa, and compliance units are Nm. An alternative set of consistent units are ton for mass, mm for length, and second for time. Then, force units are Newton, stress units are MPa, and compliance units are Nmm. It should be carefully noted that *ld*, *nelx*, *nely*, and *rmin* are in units of element, regardless of the element edge length *L*. That is, an *ld* value of 3 means that load is distributed on 3 elements. Additionally, *xlim* and *vlim* have normalized values between 0 and 1. That is, a *vlim* value of 0.5 means that 50% of the material amount of a full solid design *(number of elements in x)* x *(number of elements in y)* is to be filled in.

The computer programs are verified against the ANSYS commercial FE software by means of comparing displacement, compliance, and stress values. It is noteworthy that the stress values presented in this work and by the computer programs are actual stress values meaning that they are not normalized, multiplied by density, or norms of actual stresses values.

**Numerical examples**

Results section consists of three parts. The first part shows that PTOs and PTOc work well for topology optimization. The second part compares PTOc to Top88, and the third compares PTOs to PTOc. In all parts, three numerical examples that are defined in Figure 2 are considered.

In all three examples, material properties are input as 1 for Young's modulus $E_0$, 0.3 for Poisson's ratio $v$, and $10^{-9}$ for Young's modulus assigned to void regions $E_{min}$. Penalty value for modified SIMP approach *penal* is set to 3. A load value of 1 (*lv*) is imposed over 3 elements (*ld*). Lower and upper bounds *xlim* on elemental density are limited to 0 and 1. Element edge length *L* and filter radius *rmin* are set to 1 and 1.5, respectively. Thickness of the domain is assumed to be equal to 1. As stated earlier, $q$ is tuned to 2 for PTOs and $\alpha$ is tuned to 0.5 for PTOc.

In the first example, right half of the MBB beam is discretized by 120x40 (*nelx* x *nely*) elements. The beam is fixed in *x*-dimension on the left edge due to symmetry and fixed in *y*-dimension on the lower-right corner. A normal force is applied on the upper-left corner. In the second example, the cantilever beam is discretized by 120x60 (*nelx* x *nely*) elements. The beam is fixed in both *x* and *y*-dimensions on the left edge and a shear force is applied at the middle of the right edge. In the third example, the L bracket is discretized by 100x40 (*nell* x *nels*) elements in long (*l*) and short (*s*) edges. The bracket is fixed in both *x* and *y*-dimensions on the upper edge and a normal force is applied on the top of the most right edge.

The first part of the results section runs PTOc and PTOs for the three examples. Initially, PTOc is run for a volume fraction 0.35 and then the output stress value is input to the PTOs as a constraint. For instance, PTOc is called to solve the MBB example by the following command

$PTOc\_mbb(0.5, 1, 1e-9, 1, 1, 3, 120, 40, 0.3, 3, 1.5, 0.35, [0,1])$

The simulation ends with a stress 1.08. Then, PTOs is called with this stress value by the following command

$PTOs\_mbb(1, 1e-9, 1, 1, 3, 120, 40, 0.3, 3, 2, 1.5, 1.08, [0,1])$

This routine is repeated for the cantilever beam and L bracket examples. The simulations converge with the results tabulated in Table 1 to the topologies shown in Figure 5. Some remarks are in order.

All six cases show that for the same stress level, PTOs results with higher compliance but lower volume fraction. On average, PTOc solutions have 6.3% less compliance; and, PTOs solutions have 7.3% less volume. The topologies are almost identical for the cantilever beam example, but they are considerably different for the MBB beam and L bracket examples. PTOc tends to have thicker structural members while PTOs inclines towards more number of structural members. The contrasts of topologies are investigated by an index defined as

$$\text{Contrast index} = \frac{NoE \text{ with } \rho_i < 0.01 \text{ or } \rho_i > 0.99}{\text{Total } NoE} \quad (15)$$

where *NoE* is the number of elements and $\rho_i$ is the elemental density. The results are given in Table 1. On average, PTOc and PTOs topologies result with 0.83 and 0.85 contrast indices, respectively. The contrast indices indicate that both PTOs and PTOc provide with near black-and-white solutions.

The user has a few options to get completely black-and-white solutions at the end of the simulation. Among these are continuation methods that suggests progressive decrease of the filter radius (Rozvany 2009) or increase of the SIMP penalization factor (Sigmund 2011) during the course of the simulation. Another option is to use post-processing tools, such as projection schemes, to drive the simulation result to a black-and-white final result (Sigmund 2011). These methods are considered to be efficient and effective, but partially heuristic.

**Table 1:** Number of iterations, volume fraction, compliance, maximum stress, and contrast index obtained from MBB beam, cantilever beam, and L bracket solved by PTOc and PTOs.

|  |  | **Number of iterations** | **Volume fraction** | **Compliance** | **Max stress** | **Contrast index** |
|---|---|---|---|---|---|---|
| **MBB beam** | PTOc | 170 | 0.35 | 266.61 | 1.08 | 0.80 |
|  | PTOs | 206 | 0.31 | 294.92 | 1.08 | 0.83 |
| **Cantilever beam** | PTOc | 106 | 0.35 | 88.54 | 0.57 | 0.85 |
|  | PTOs | 164 | 0.34 | 90.62 | 0.57 | 0.88 |
| **L bracket** | PTOc | 78 | 0.35 | 235.25 | 1.05 | 0.83 |
|  | PTOs | 187 | 0.33 | 248.97 | 1.05 | 0.85 |

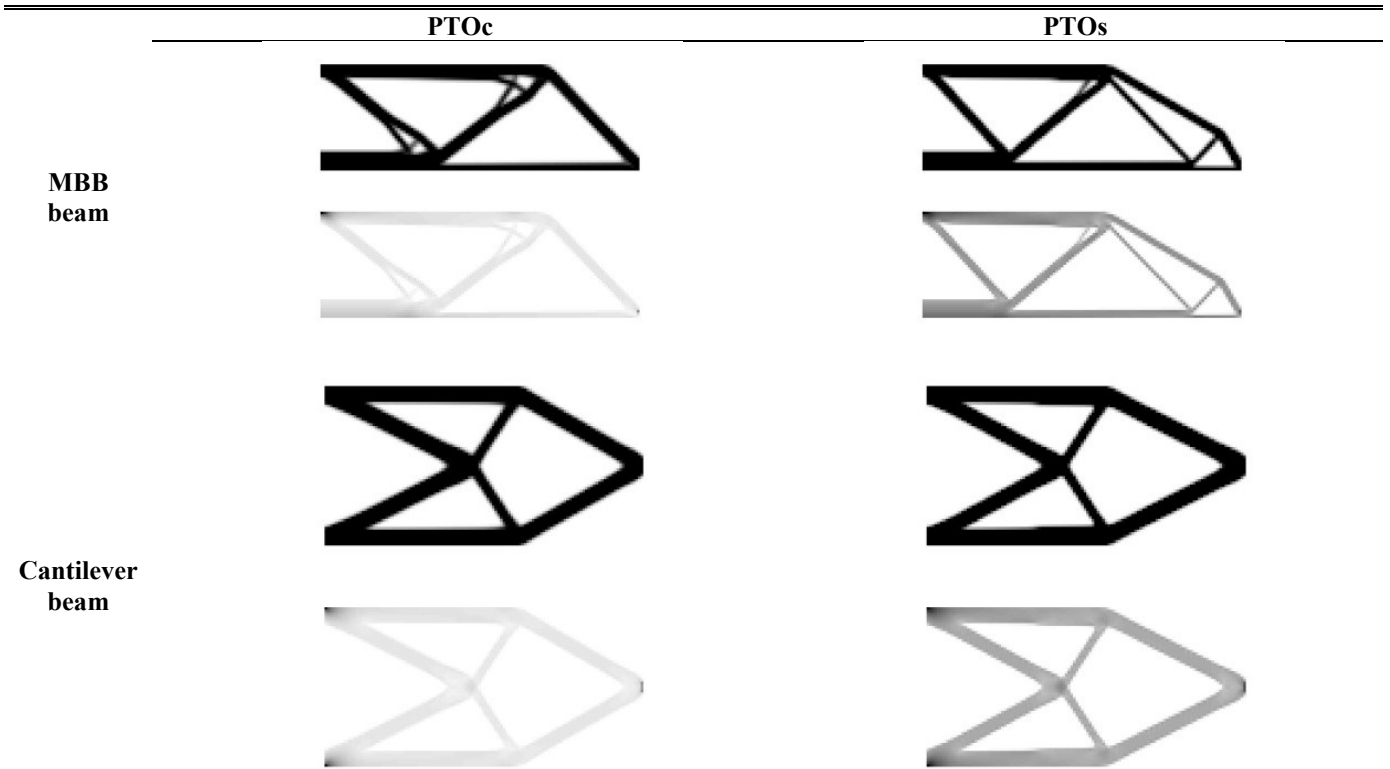

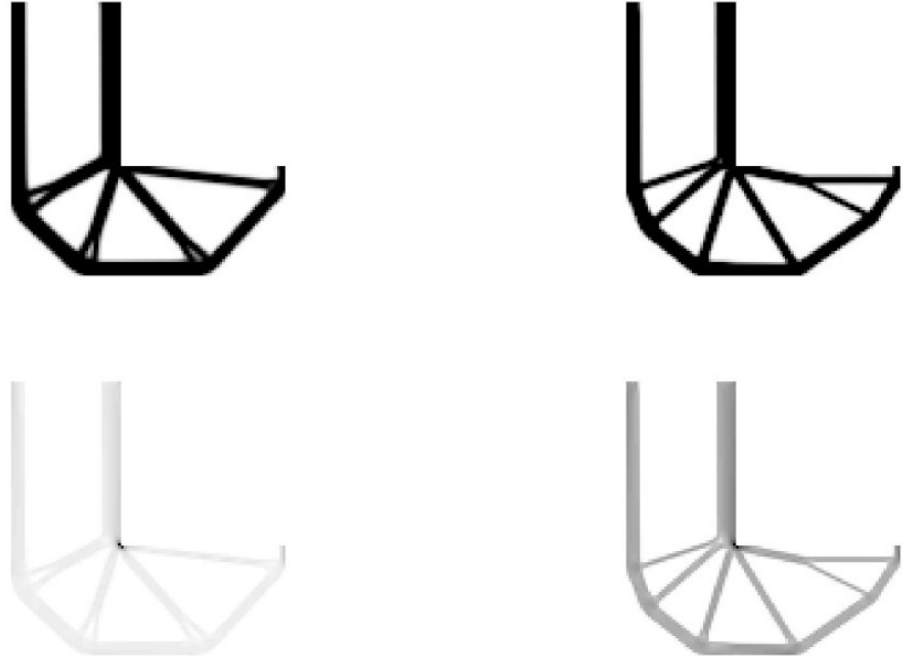

**L bracket**

**Figure 5:** Topologies and compliance (PTOc) or stress (PTOs) distributions obtained from the MBB beam, cantilever beam, and L bracket examples.

The second part of results section compares PTOc to Top88 for the three examples. It should be clarified that the original Top88 code is only for the MBB beam example; but, it has been extended to solve the cantilever beam and L bracket examples. In this connection, Top88 represents an OC method with gradients. For each example, both programs are called by a set of identical inputs. For instance, material properties and penalization factor, density filter and its radius, and loading value and distribution are set the same. As a result, simulations for each method are identical except the optimization algorithms. PTOc and Top88 are run for a number of volume fractions *vlim* from 0.25 to 0.50 in increments of 0.05. Figure 6 shows comparison of compliances for the three examples. The figures indicate that the compliance versus volume fraction curves of PTOc and Top88 are indistinguishable for all three examples. Also, average number of iterations is compared. In this regard, PTOc takes 26.6% more, 0.4% less, and 23.7% less iterations than Top88 for the examples in the presented order. This proves that none of the methods is superior to the other in terms of efficiency in general but they have varying performances depending on the example.

Figure 7 compares topologies obtained by running PTOc and Top88 for a volume fraction 0.35 for three examples. Topologies are similar for the cantilever beam example and remarkably different for the MBB beam and L bracket examples. The most prominent difference is the tiny feature near the loading in the L bracket topology solved by PTOc. Such a tiny feature is not a good design practice since it is fragile against loadings in traverse directions. Thus, these kinds of considerations should be made by the designer in the post-processing phase. The topologies are also compared by their contrast indices. Contrast indices for Top88 topologies are 0.81, 0.86, and 0.83 for the examples in the presented order. Compared to contrast indices of PTOc in Table 1, contrast indices between the two methods are not different more than 0.01.

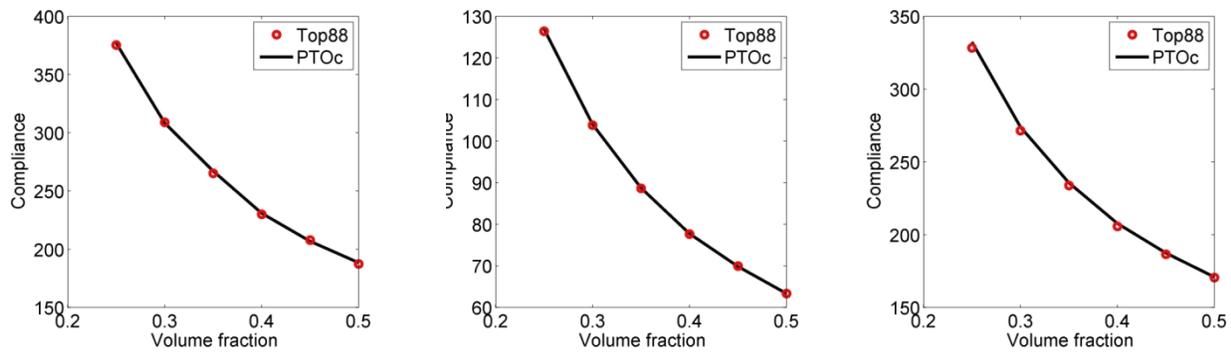

**Figure 6:** Comparison of compliance versus volume fraction curves of PTOc and Top88 for the MBB beam (left), cantilever beam (center), and L bracket (right) examples.

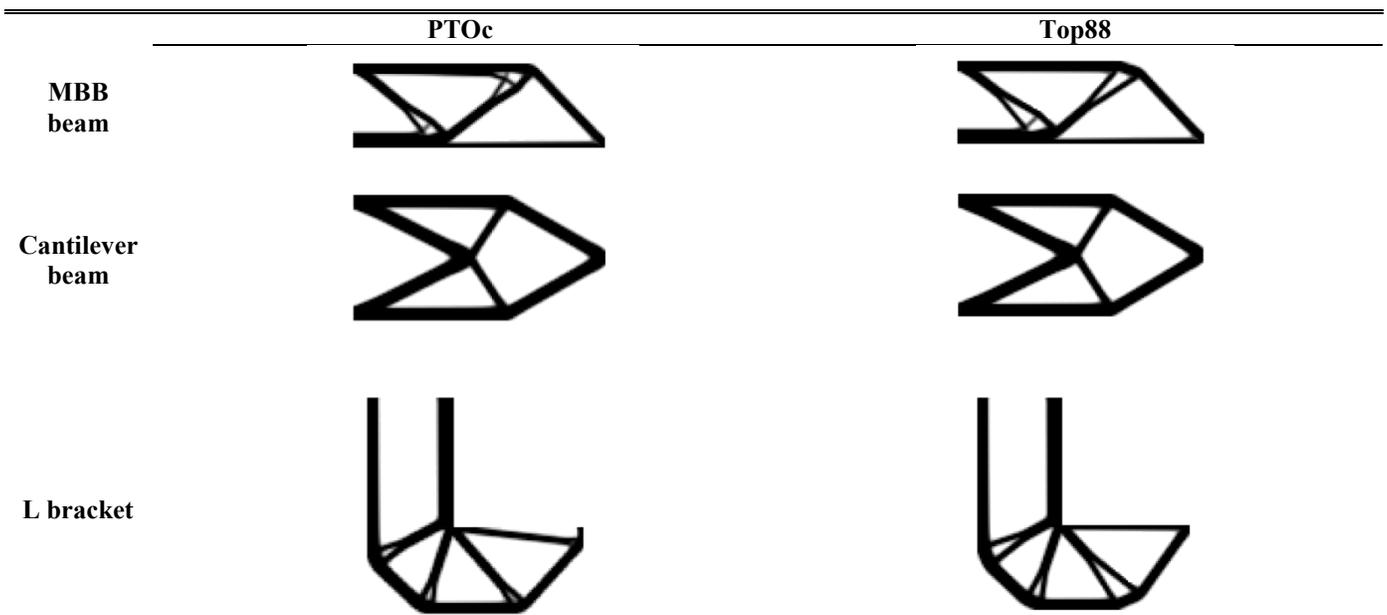

**Figure 7:** Comparison of topologies of PTOc and Top88 for the MBB beam, cantilever beam, and L bracket examples.

The third part of results section compares PTOs to PTOc. The comparison is conducted iteratively starting from PTOc at 0.5 volume fraction. The output stress of PTOc is then input to the PTOs. Following, the output volume fraction of PTOs is input back to the PTOc, and so on. Figure 8 shows the results for MBB beam, cantilever beam, and L bracket examples. The figures show that PTOs performs better than PTOc by means of providing less volume fraction for the same level of stress and less stress for the same level of volume fraction for all three examples. This improvement is more pronounced in the MBB beam example compared to other two examples. The results are also quantified by taking the average improvements for each example, see Table 2. The results prove that the extent of improvements depend on the example. On average, though, PTOs provides 8.4% less stress for the same level of volume fraction and 5.9% less volume fraction for the same level of stress.

**Table 2:** Quantitative comparison of stress and compliance for PTOs and PTOc.

|  | PTOs improvement of stress (%) | PTOs improvement of volume fraction (%) |
|---|---|---|
| MBB beam | 12.8 | 9.5 |
| Cantilever beam | 5.5 | 4.1 |
| L bracket | 7.0 | 4.0 |
| Average | 8.4 | 5.9 |

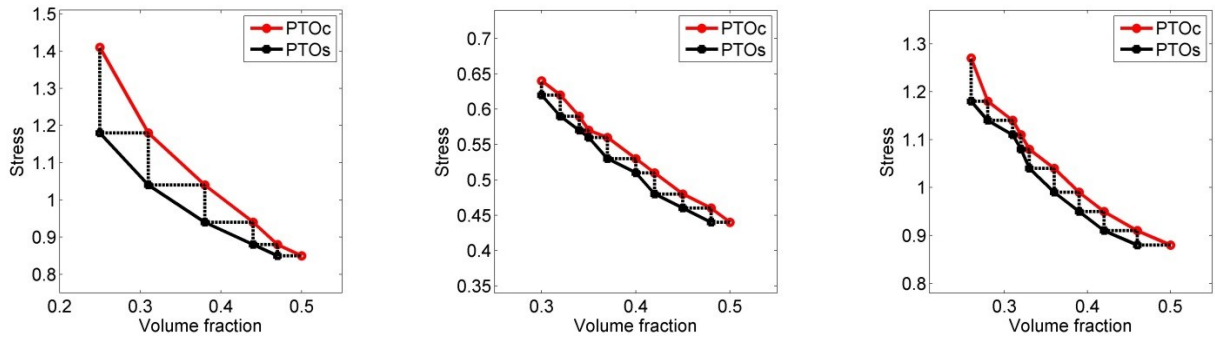

**Figure 8:** Comparison of stress versus volume fraction curves of PTOs and PTOc for the MBB beam (left), cantilever beam (center), and L bracket (right) examples. Dashed lines indicate the links between PTOs and PTOc. A horizontal dashed line means stress output of PTOc is input to the PTOs and a horizontal dashed line means volume fraction output of PTOs is input to PTOc.

## Conclusions

A new topology optimization method, named PTO, is introduced. It is a non-gradient method, and thus eliminates difficulties emerged from analytical derivations and computational implementation of gradients. The achieved balance comes with a price of weaker mathematical rigor but worthy simplicity at the same time. The method possesses considerable efficiency and accuracy considering its simplicity. Even more, the various comparisons to results generated by the Top88 code show that PTOc attains very similar results without use of gradients while maintaining same level of efficiency. On the other hand, although it is not presented here, PTOs has always been thought to be not as efficient and accurate as the state of the art methods of topology optimization field that solves stress problems for continua, especially the ones utilizing gradients (Sigmund 2011). A comparison is left for future work.

PTO can be useful especially in educational and industrial purposes owing to its simplicity. As pointed out by Rozvany (2009), industrial practitioners tend to work with methods that are easier to understand and manipulate. Naturally, students and newcomers to the topology optimization field share alike manners (Sigmund 2001). The method can also be useful in research due to its flexibility and extensibility. For the above purposes, two computer programs that solve the MBB beam example for stress and compliance problems are presented. The programs are individually coded in MATLAB as standalone functions and they are publicly shared in the website www.ptomethod.org. The website will be maintained with new versions, publications, extensions, and other up-to-date information.

There is more room to investigate and enhance the method, but they are left for future work. First of all, a more comprehensive parametric work is required to utilize the method at its best. Second, mesh dependency of the method is to be investigated more carefully. It is argued that filtering leads to mesh independent solutions, but this point of view is only supported by comparison of topologies (Andreassen et al. 2011). The authors believe that quantitative comparisons should be carried out alongside. Third, it should be investigated whether the method would benefit from clustering of elements so that the constraints could be imposed on these clusters. It was shown that clustering of elements yield more efficient results (Le et al. 2010). In the current work, the method considers only one cluster that includes the whole domain. The listed future works are subject to ongoing research and will be presented in an upcoming paper.

## Acknowledgements

The financial support from Mascaro Center for Sustainable Innovations (MCSI) at University of Pittsburgh is gratefully acknowledged.

## Appendix A – PTOs

```matlab
% Proportional Topology Optimization stress (PTOs) - Half MBB Beam - (2014)
%%%%%%%%%%%%%%%%%%%%%%%%%%%%%%%%%%%%%%%%%%%%%%%%%%%%%%%%%%%%%%%%%%%%%%%%%%%%
function x = PTOs_mbb(E0,Emin,L,lv,ld,nelx,nely,nu,penal,q,rmin,vmslim,xlim)
% Setup Finite Element Analysis
A11 = [12  3 -6 -3;  3 12  3  0; -6  3 12 -3; -3  0 -3 12];
A12 = [-6 -3  0  3; -3 -6 -3 -6;  0 -3 -6  3;  3 -6  3 -6];
B11 = [-4  3 -2  9;  3 -4 -9  4; -2 -9 -4 -3;  9  4 -3 -4];
B12 = [ 2 -3  4 -9; -3  2  9 -2;  4  9  2  3; -9 -2  3  2];
KE = 1/(1-nu^2)/24*([A11 A12;A12' A11]+nu*[B11 B12;B12' B11]);
nodenrs = reshape(1:(1+nelx)*(1+nely),1+nely,1+nelx);
edofVec = reshape(2*nodenrs(1:end-1,1:end-1)+1,nelx*nely,1);
edofMat = repmat(edofVec,1,8)+repmat([0 1 2*nely+[2 3 0 1] -2 -1],nelx*nely,1);
iK = reshape(kron(edofMat,ones(8,1))',64*nelx*nely,1);
jK = reshape(kron(edofMat,ones(1,8))',64*nelx*nely,1);
% Define Loads and Supports
iF = 2*(nely+1)*(0:ld-1)+2;
jF = ones(1,ld);
sF = -lv/ld*ones(ld,1);
F = sparse(iF,jF,sF,2*(nely+1)*(nelx+1),1);
% Define Displacement and DOF Sets
U = zeros(2*(nely+1)*(nelx+1),1);
fixeddofs = union(1:2:2*(nely+1),2*((nelx+1)*(nely+1)-ld+1:(nelx+1)*(nely+1)));
alldofs = 1:2*(nely+1)*(nelx+1);
freedofs = setdiff(alldofs,fixeddofs);
% Setup Stress Analysis
B = (1/2/L)*[-1 0 1 0 1 0 -1 0; 0 -1 0 -1 0 1 0 1; -1 -1 -1 1 1 1 1 -1];
DE = (1/(1-nu^2))*[1 nu 0; nu 1 0; 0 0 (1-nu)/2];
% Setup Filter
iW = ones(nelx*nely*(2*(ceil(rmin)-1)+1)^2,1);
jW = ones(size(iW));
sW = zeros(size(iW));
k = 0;
for i1 = 1:nelx
  for j1 = 1:nely
    e1 = (i1-1)*nely+j1;
    for i2 = max(i1-(ceil(rmin)-1),1):min(i1+(ceil(rmin)-1),nelx)
      for j2 = max(j1-(ceil(rmin)-1),1):min(j1+(ceil(rmin)-1),nely)
        e2 = (i2-1)*nely+j2;
        k = k+1;
        iW(k) = e1;
        jW(k) = e2;
        sW(k) = max(0,rmin-sqrt((i1-i2)^2+(j1-j2)^2));
      end
    end
  end
end
w = sparse(iW,jW,sW);
W = bsxfun(@rdivide,w,sum(w,2));
% Initialize Iteration
x = repmat(0.5,nely,nelx);
loop = 0;
% Run Iteration
while (true)
 loop = loop+1;
 % Finite Element Analysis
 E = Emin+x(:)'.^penal*(E0-Emin);
 sK = reshape(KE(:)*E,64*nelx*nely,1);
 K = sparse(iK,jK,sK); K = (K+K')/2;
 U(freedofs) = K(freedofs,freedofs)\F(freedofs);
```

```matlab
% Stress Calculation
s = (U(edofMat)*(DE*B)').*repmat(E',1,3);
vms = reshape(sqrt(sum(s.^2,2)-s(:,1).*s(:,2)+2.*s(:,3).^2),nely,nelx);
% Compliance Calculation
ce = E'.*sum((U(edofMat)*KE).*U(edofMat),2);
C = reshape(ce,nely,nelx);
% Print Results
fprintf('It:%5i Max_vms:%5.2f Comp:%8.2f Vol:%5.2f Res:%6.3f\n',...
        loop,max(vms(:)),sum(C(:)),mean(x(:)),abs(max(vms(:))-vmslim));
% Plot Results
colormap(flipud(gray));
subplot(2,1,1); imagesc(x); axis equal off; text(2,-2,'x');
subplot(2,1,2); imagesc(vms); axis equal off; text(2,-2,'vms'); drawnow;
% Check Stop Criteria
if (abs(max(vms(:))-vmslim) < 0.001 && loop > 50); break; end;
% Optimization Algorithm (PTOs)
if (max(vms(:)) > vmslim)
  xTarget = sum(x(:))+0.001*numel(x);
else
  xTarget = sum(x(:))-0.001*numel(x);
end
xRemaining = xTarget;
x(:) = 0;
vms_proportion = vms.^q/sum(sum(vms.^q));
while (xRemaining > 0.001)
 xDist = xRemaining.*vms_proportion;
 x(:) = x(:)+W*xDist(:);
 x = max(min(x,xlim(2)),xlim(1));
 xRemaining = xTarget-sum(x(:));
 end
end
end
%
%%%%%%%%%%%%%%%%%%%%%%%%%%%%%%%%%%%%%%%%%%%%%%%%%%%%%%%%%%%%%%%%%%%%%%%%%%%
% Copyright (C) 2014 University of Pittsburgh. All rights reserved.
%
% Any person who obtained a copy of this software can (in part or whole) copy,
% modify, merge, publish, and distribute the software on condition of retaining
% this license with the software. The user is allowed to utilize the software
% for all purposes but commercial. Also, appropriate credit must be provided.
%
% The software is provided "as is", without warranty of any kind, express or
% implied, including but not limited to the warranties of merchantability,
% fitness for a particular purpose and noninfringement. In no event shall the
% authors or copyright holders be liable for any claim, damage or other
% liability, whether in an action of contract, tort or otherwise, arising from,
% out of or in connection with the software or the use or other dealing in the
% software.
%
% The software is coded by Emre Biyikli (biyikli.emre@gmail.com) and Albert C.
% To (albertto@pitt.edu). The software is substantially inherited from
% Andreassen E, et al. "Efficient topology optimization in MATLAB using 88 lines
% of code." Structural and Multidisciplinary Optimization 43.1 (2011): 1-16.
% The software can be downloaded from www.ptomethod.org.
% The journal article to the software is ...
%
```

## Appendix B – PTOc

```matlab
% Proportional Topology Optimization compliance (PTOc) - Half MBB beam - (2014)
%%%%%%%%%%%%%%%%%%%%%%%%%%%%%%%%%%%%%%%%%%%%%%%%%%%%%%%%%%%%%%%%%%%%%%%%%%%%%
function x = PTOc_mbb(alpha,E0,Emin,L,lv,ld,nelx,nely,nu,penal,rmin,vlim,xlim)
% Setup Finite Element Analysis
A11 = [12  3 -6 -3;  3 12  3  0; -6  3 12 -3; -3  0 -3 12];
A12 = [-6 -3  0  3; -3 -6 -3 -6;  0 -3 -6  3;  3 -6  3 -6];
B11 = [-4  3 -2  9;  3 -4 -9  4; -2 -9 -4 -3;  9  4 -3 -4];
B12 = [ 2 -3  4 -9; -3  2  9 -2;  4  9  2  3; -9 -2  3  2];
KE = 1/(1-nu^2)/24*([A11 A12;A12' A11]+nu*[B11 B12;B12' B11]);
nodenrs = reshape(1:(1+nelx)*(1+nely),1+nely,1+nelx);
edofVec = reshape(2*nodenrs(1:end-1,1:end-1)+1,nelx*nely,1);
edofMat = repmat(edofVec,1,8)+repmat([0 1 2*nely+[2 3 0 1] -2 -1],nelx*nely,1);
iK = reshape(kron(edofMat,ones(8,1))',64*nelx*nely,1);
jK = reshape(kron(edofMat,ones(1,8))',64*nelx*nely,1);
% Define Loads and Supports
iF = 2*(nely+1)*(0:ld-1)+2;
jF = ones(1,ld);
sF = -lv/ld*ones(ld,1);
F = sparse(iF,jF,sF,2*(nely+1)*(nelx+1),1);
% Define Displacement and DOF Sets
U = zeros(2*(nely+1)*(nelx+1),1);
fixeddofs = union(1:2:2*(nely+1),2*((nelx+1)*(nely+1)-ld+1:(nelx+1)*(nely+1)));
alldofs = 1:2*(nely+1)*(nelx+1);
freedofs = setdiff(alldofs,fixeddofs);
% Setup Stress Analysis
B = (1/2/L)*[-1 0 1 0 1 0 -1 0; 0 -1 0 -1 0 1 0 1; -1 -1 -1 1 1 1 1 -1];
DE = (1/(1-nu^2))*[1 nu 0; nu 1 0; 0 0 (1-nu)/2];
% Setup Filter
iW = ones(nelx*nely*(2*(ceil(rmin)-1)+1)^2,1);
jW = ones(size(iW));
sW = zeros(size(iW));
k = 0;
for i1 = 1:nelx
  for j1 = 1:nely
    e1 = (i1-1)*nely+j1;
    for i2 = max(i1-(ceil(rmin)-1),1):min(i1+(ceil(rmin)-1),nelx)
      for j2 = max(j1-(ceil(rmin)-1),1):min(j1+(ceil(rmin)-1),nely)
        e2 = (i2-1)*nely+j2;
        k = k+1;
        iW(k) = e1;
        jW(k) = e2;
        sW(k) = max(0,rmin-sqrt((i1-i2)^2+(j1-j2)^2));
      end
    end
  end
end
w = sparse(iW,jW,sW);
W = bsxfun(@rdivide,w,sum(w,2));
% Initialize Iteration
x = repmat(vlim,nely,nelx);
xNew = zeros(size(x));
loop = 0;
change = Inf;
% Run Iteration
while (true)
 loop = loop+1;
 % Finite Element Analysis
```

```matlab
  E = Emin+x(:)'.^penal*(E0-Emin);
  sK = reshape(KE(:)*E,64*nelx*nely,1);
  K = sparse(iK,jK,sK); K = (K+K')/2;
  U(freedofs) = K(freedofs,freedofs)\F(freedofs);
  % Stress Calculation
  s = (U(edofMat)*(DE*B)').*repmat(E',1,3);
  vms = reshape(sqrt(sum(s.^2,2)-s(:,1).*s(:,2)+2.*s(:,3).^2),nely,nelx);
  % Compliance Calculation
  ce = E'.*sum((U(edofMat)*KE).*U(edofMat),2);
  C = reshape(ce,nely,nelx);
  % Print Results
  fprintf('It:%5i Max_vms:%5.2f Comp:%8.2f Vol:%5.2f Ch:%6.3f\n',...
          loop,max(vms(:)),sum(C(:)),mean(x(:)),change);
  % Plot Results
  colormap(flipud(gray));
  subplot(2,1,1); imagesc(x); axis equal off; text(2,-2,'x');
  subplot(2,1,2); imagesc(C); axis equal off; text(2,-2,'C'); drawnow;
  % Check Stop Criteria
  if(change < 0.01 && loop > 50); break; end;
  % Optimization Algorithm (PTOc)
  xTarget = nelx*nely*vlim;
  xRemaining = xTarget;
  xNew(:) = 0;
  C_proportion = C/sum(C(:));
  while (xRemaining > 0.001)
   xDist = xRemaining.*C_proportion;
   xNew(:) = xNew(:)+W*xDist(:);
   xNew = max(min(xNew,xlim(2)),xlim(1));
   xRemaining = xTarget-sum(xNew(:));
  end
  x = alpha*x+(1-alpha)*xNew;
  change = max(abs((1/alpha-1)*(xNew(:)-x(:))));
 end
end
%
%%%%%%%%%%%%%%%%%%%%%%%%%%%%%%%%%%%%%%%%%%%%%%%%%%%%%%%%%%%%%%%%%%%%%%%%%%%
% Copyright (C) 2014 University of Pittsburgh. All rights reserved.
%
% Any person who obtained a copy of this software can (in part or whole) copy,
% modify, merge, publish, and distribute the software on condition of retaining
% this license with the software. The user is allowed to utilize the software
% for all purposes but commercial. Also, appropriate credit must be provided.
%
% The software is provided "as is", without warranty of any kind, express or
% implied, including but not limited to the warranties of merchantability,
% fitness for a particular purpose and noninfringement. In no event shall the
% authors or copyright holders be liable for any claim, damage or other
% liability, whether in an action of contract, tort or otherwise, arising from,
% out of or in connection with the software or the use or other dealing in the
% software.
%
% The software is coded by Emre Biyikli (biyikli.emre@gmail.com) and Albert C.
% To (albertto@pitt.edu). The software is substantially inherited from
% Andreassen E, et al. "Efficient topology optimization in MATLAB using 88 lines
% of code." Structural and Multidisciplinary Optimization 43.1 (2011): 1-16.
% The software can be downloaded from www.ptomethod.org.
% The journal article to the software is ...
%
```